\documentclass[pra,aps,twocolumn,showpacs]{revtex4}
\usepackage{epsf}
\usepackage{amsmath}
\usepackage{amssymb}

\begin{document}

\title{Ordering two-qubit states with concurrence and negativity}

\author{Adam Miranowicz and Andrzej Grudka}
\date{\today}

\affiliation{Faculty of Physics, Adam Mickiewicz University, 61-614
Pozna\'n, Poland}

\begin{abstract}
We study the ordering of two-qubit states with respect to the
degree of bipartite entanglement using the Wootters concurrence --
a measure of the entanglement of formation, and the negativity -- a
measure of the entanglement cost under the
positive-partial-transpose-preserving operations. For two-qubit
pure states, the negativity is the same as the concurrence.
However, we demonstrate analytically on simple examples of various
mixtures of Bell and separable states that the entanglement
measures can impose different orderings on the states. We show
which states, in general, give the maximally different predictions,
(i) when one of the states has the concurrence greater but the
negativity smaller than those for the other state, and (ii) when
the states are entangled to the same degree according to one of the
measures, but differently according to the other.

\end{abstract}

\pacs{03.65.Ud, 42.50.Dv}

\maketitle

\pagenumbering{arabic}

The entropy of entanglement is essentially the unique measure of
entanglement for pure states of bipartite systems \cite{popescu}.
By contrast, a generalization of the entropy of entanglement to
describe mixed states is by no means unique (even for two qubits)
leading to various entanglement measures (for a review see
\cite{horodecki}), including entanglement of formation, distillable
entanglement \cite{bennett1,bennett2}, and relative entropy of
entanglement \cite{vedral}. The quantification of mixed-state
entanglement is still in early stages with many open questions.

Eisert and Plenio \cite{eisert} raised an intriguing problem of
ordering the density operators with respect to the amount of
entanglement. Specifically, certain two entanglement measures $E'$
and $E''$ are defined to give the same state ordering if the
condition \cite{eisert}
\begin{equation}
E'(\rho_1) < E'(\rho_2) \Leftrightarrow E''(\rho_1) < E''(\rho_2)
\label{N01}
\end{equation}
is satisfied for any density operators $\rho_1$ and $\rho_2$. No
counterexample to Eq. (\ref{N01}) can be found by comparing pure
states only, or pure and Werner states \cite{eisert}. However, the
standard entanglement measures do not give the same ordering in the
sets of two-qubit mixed states, as first observed by applying the
Monte Carlo simulation by Eisert and Plenio \cite{eisert}, and then
investigated by others in Refs.
\cite{zyczkowski99,virmani,zyczkowski02,wei1,wei2,miran}.
Counterexamples to Eq. (\ref{N01}) can also be constructed for
$d$-level qudit pure states if $d\ge 3$ as shown by \.Zyczkowski
and Bengtsson \cite{zyczkowski02}. Virmani and Plenio
\cite{virmani} proved that all good asymptotic entanglement
measures, which reduce to the entropy of entanglement for pure
states, are either equivalent or do not have the same state
ordering.   The property that ordering of some states depends on
the applied measures of entanglement ``in itself is a very
surprising conclusion'' \cite{virmani} but is physically reasonable
as these incomparable states cannot be transformed to each other
with unit efficiency by any local operations and classical
communication (LOCC).

In this paper, we present an analysis of different orderings of
two-qubit states induced by concurrence and negativity. The
ordering of two-qubit states by the same measures has already been
studied by Eisert and Plenio \cite{eisert}, but by using only a
numerical simulation (of $10^4$ pairs of entangled states
\{$\rho_1,\rho_2$\}). For three qubits, analytical counterexamples
to (\ref{N01}) are known even for pure states
\cite{zyczkowski02,wei2}.

First, we briefly describe the entanglement measures important for
our comparison.

The entanglement of formation $E_{F}$ of a mixed state $\rho$,
according to Bennett et al. \cite{bennett1,bennett2}, is the
minimized average entanglement of any ensemble of pure states
$|\psi _{i}\rangle$ realizing $\rho$:
\begin{equation}
E_{F}({\rho })=\inf \sum_{i}p_{i}E(|\psi _{i}\rangle \langle \psi
_{i}|) \label{N02}
\end{equation}
where infimum is taken over all pure-state decompositions $\rho
=\sum_{i}p_{i}|\psi _{i}\rangle \langle \psi _{i}|$ and  $E(|\psi
_{i}\rangle \langle \psi _{i}|)$ is the entropy of entanglement
easily determined by the von Neumann entropy. In a special case of
two qubits, Wootters \cite{wootters} proved that the entanglement
of formation of a state ${\rho}$ is given by a simple formula
\begin{equation}
E_{F}({\rho })=H\left(\frac{1}{2}[1+\sqrt{1-C^{2}(\rho)}]\right)
\label{N03}
\end{equation}
where $H(x)=-x\log _{2}x-(1-x)\log _{2}(1-x)$ is the binary entropy
with the argument related to the Wootters concurrence defined by
\begin{equation}
C({\rho })=\max \{0,\lambda_1-\lambda_2-\lambda_3-\lambda_4\}
\label{N04}
\end{equation}
where the $\lambda _{i}$'s  are (in nonincreasing order) the square
roots of the eigenvalues of ${\rho }({\sigma }_{y}\otimes {\sigma
}_{y}){\rho }^{\ast }({ \sigma }_{y}\otimes {\sigma }_{y})$;
${\sigma }_{y}$ is the Pauli spin matrix and complex conjugation is
denoted by asterisk. Both $E_{F}(\rho)$ and $C(\rho)$ range from 0
for a separable state to 1 for a maximally entangled state.

We also consider another entanglement measure referred to as the
negativity, which can be considered a quantitative version of the
Peres-Horodecki criterion \cite{peres}. The negativity for a
two-qubit state $\rho$ can be defined as
\cite{zyczkowski98,eisert,vidal}:
\begin{equation}
{N}({\rho })=\max \{0,-2\mu _{\min}\}  \label{N05}
\end{equation}
where $\mu _{\min}$ is the minimal eigenvalue of the partial
transpose of $\rho$. Similarly to the concurrence, the negativity,
given by (\ref{N05}), ranges from 0 for a separable state to 1 for
a maximally entangled state. As shown by Vidal and Werner
\cite{vidal}, the negativity is an entanglement monotone (including
convexity) thus can be considered a useful measure of entanglement.
The logarithmic negativity defined by \cite{vidal}
\begin{equation}
E_N(\rho)=\log _{2}[{N} ({\rho })+1], \label{N06}
\end{equation}
measures the entanglement cost of a quantum state $\rho$ for the
exact preparation of any finite number of copies of the state under
quantum operations preserving the positivity of the partial
transpose (PPT) as proposed by Audenaert et al. \cite{audenaert}
and proved by Ishizaka \cite{ishizaka} for any two-qubit states.
Moreover, the logarithmic negativity determines upper bounds on the
teleportation capacity and the entanglement of distillation
\cite{vidal}.

In the following, we will analyze two-qubit states violating the
condition (\ref{N01}) induced by the concurrence (the entanglement
of formation) and negativity (the PPT-entanglement cost) only. And
thus by referring to (\ref{N01}) we always mean $E'=C$ and $E''=N$.

For an arbitrary two-qubit pure state, given by
\begin{equation}
|\Psi \rangle =c_{00}|00\rangle +c_{01}|01\rangle +c_{10}|10\rangle
+c_{11}|11\rangle, \label{N07}
\end{equation}
where $c_{ij}$ are the normalized complex amplitudes, the
concurrence and negativity are the same and simply given by
\begin{equation}
N(|\Psi \rangle)=C(|\Psi \rangle)=2|c_{00}c_{11}-c_{01}c_{10}|.
\label{N08}
\end{equation}
Nevertheless, $N(\rho)$ and $C(\rho)$ can differ for a mixed state
$\rho$. In general, as shown by Verstraete et al.
\cite{verstraete}, the negativity $N(\rho)$ of a two-qubit state
$\rho$ can never exceed its concurrence $C(\rho)$ and
\begin{eqnarray}
N(\rho)&\ge&\sqrt{[1-C(\rho)]^2+C^2(\rho)} -[1-C(\rho)] \label{N09}
\end{eqnarray}
as presented in figures 1 and 3. The states corresponding to these
lower and upper bounds have the minimal (maximal) negativity for a
fixed concurrence and for short we shall refer to as the MinNeg
(MaxNeg) states. The class of the MaxNeg states can be
characterized by the condition that the eigenvector corresponding
to the negative eigenvalue of the partial transpose of $\rho$ is a
Bell state \cite{verstraete}. Apart from pure states (\ref{N07}),
the class of the MaxNeg states includes the Bell diagonal states
\cite{wootters,verstraete} with the celebrated Werner states
defined by \cite{werner}
\begin{equation}
\rho_W(p) = p |\psi_B\rangle \langle \psi_B|+\frac{1-p}{4}I\otimes
I \label{N10}
\end{equation}
where the parameter $p\in\langle 0,1\rangle$; $I$ is the identity
operator of a single qubit, and $|\psi_B\rangle$ is the singlet
state
\begin{equation}
|\psi_B\rangle =\frac{1}{\sqrt{2}}(|01\rangle - |10\rangle).
\label{N11}
\end{equation}
The negativity and concurrence of $\rho_W(p)$ are equal to each
other for any value of $p$ as given by
\begin{equation}
N(\rho_W(p))=C(\rho_W(p))=\max\{0,\frac{3p-1}{2}\}. \label{N12}
\end{equation}
Since both pure and Werner two-qubit states are the MaxNeg states,
thus it is clear why they do not violate condition (\ref{N01}),
which is another explanation of the Eisert and Plenio result
\cite{eisert}.

The structure of the class of the MinNeg states was given by
Verstraete et al. \cite{verstraete} as a solution of the Lagrange
constrained problem for the manifold of states with constant
concurrence. They found that the MinNeg states have two vanishing
eigenvalues and the other two corresponding to eigenvectors which
are a Bell state and separable state orthogonal to it. As an
example of the MinNeg state, we analyze the following state
\cite{horodecki}
\begin{equation}
\rho_H(p) = p |\psi_B\rangle \langle \psi_B|+(1-p)|00\rangle
\langle 00| \label{N13}
\end{equation}
where $0\le p \le 1$ and $|\psi_B\rangle$ is the Bell state given
by (\ref{N11}). The concurrence and negativity of $\rho_H$ are
given by
\begin{eqnarray}
C(\rho_H)&=& p, \notag \\
N(\rho_H)&=&\sqrt{(1-p)^2+p^2}-(1-p), \label{N14}
\end{eqnarray}
respectively, being equal to each other for $p=0$ and $p=1$ only.
The state, given by (\ref{N13}), is a mixture of a maximally
entangled state and a separable state orthogonal to it as required
by the Verstraete et al. condition for the MinNeg states
\cite{verstraete}. Thus, by replacing $|00\rangle$ in (\ref{N13})
by another separable state orthogonal to $|\psi_B\rangle$ (e.g., by
$|11\rangle$ or $(|00\rangle+|01\rangle+|10\rangle+|11\rangle)/2$),
other MinNeg states satisfying (\ref{N14}) can be obtained.

\begin{figure}
\centerline{\epsfxsize=5cm\epsfbox{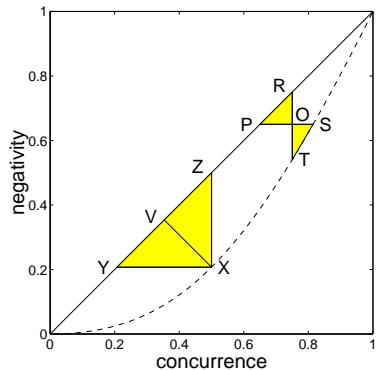}} \caption{Negativity
versus concurrence. Interpretation of the marked regions is given
in the text.}
\end{figure}
\begin{figure}
\centerline{\epsfxsize=5cm\epsfbox{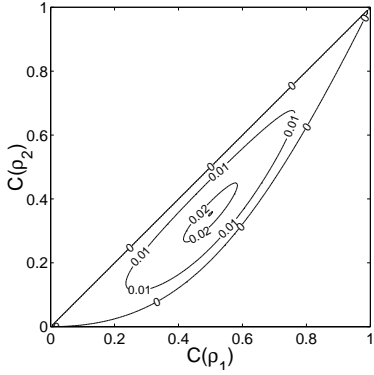}} \caption{Contour plot
of $\delta(\rho_1,\rho_2)$ vs $C(\rho_1)$ and $C(\rho_2)$ for the
MaxNeg states $\rho_1$ and the MinNeg states $\rho_2$.}
\end{figure}
\begin{figure}
\centerline{\epsfxsize=3.0cm\epsfbox{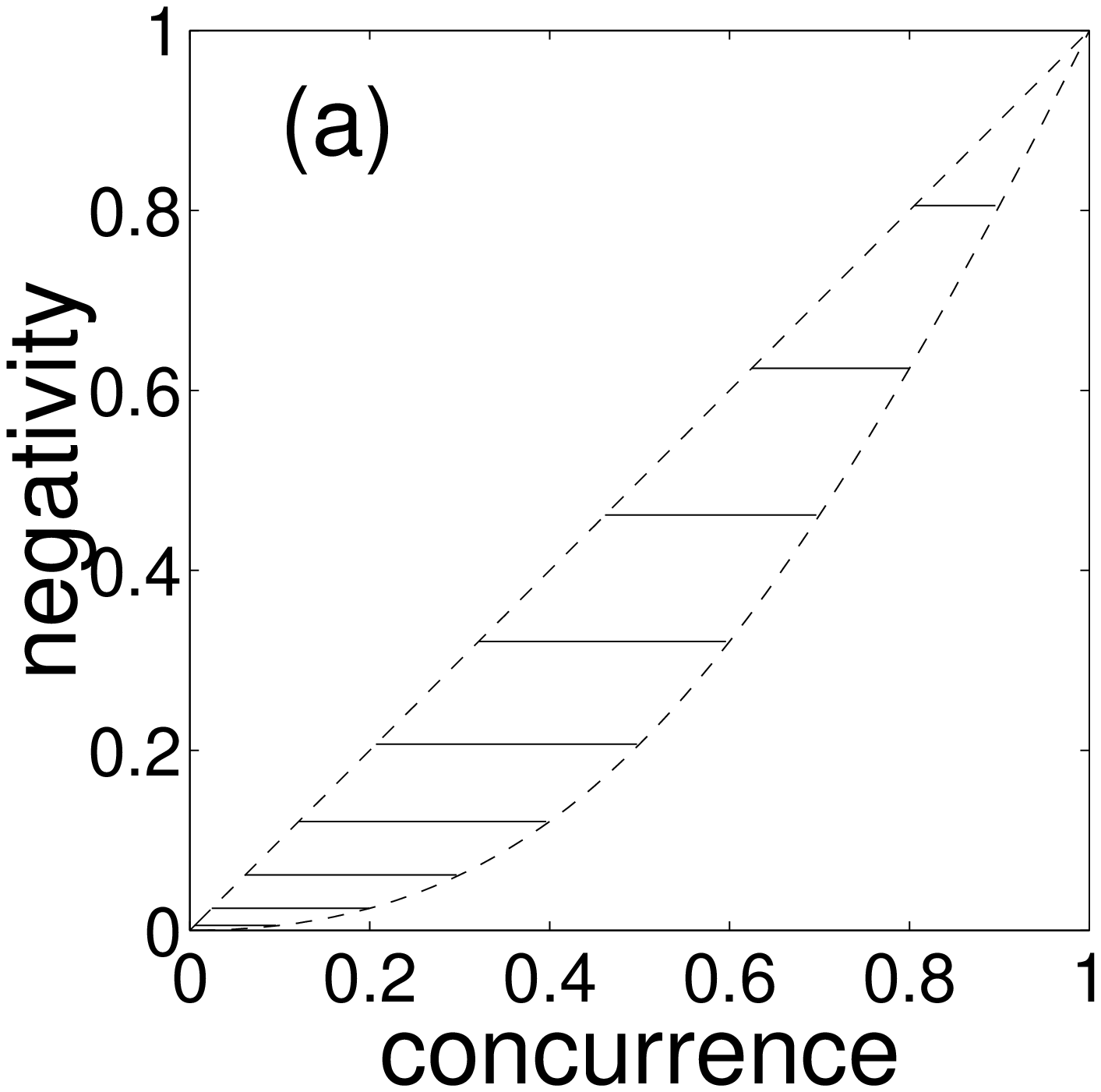}
\epsfxsize=2.8cm\epsfbox{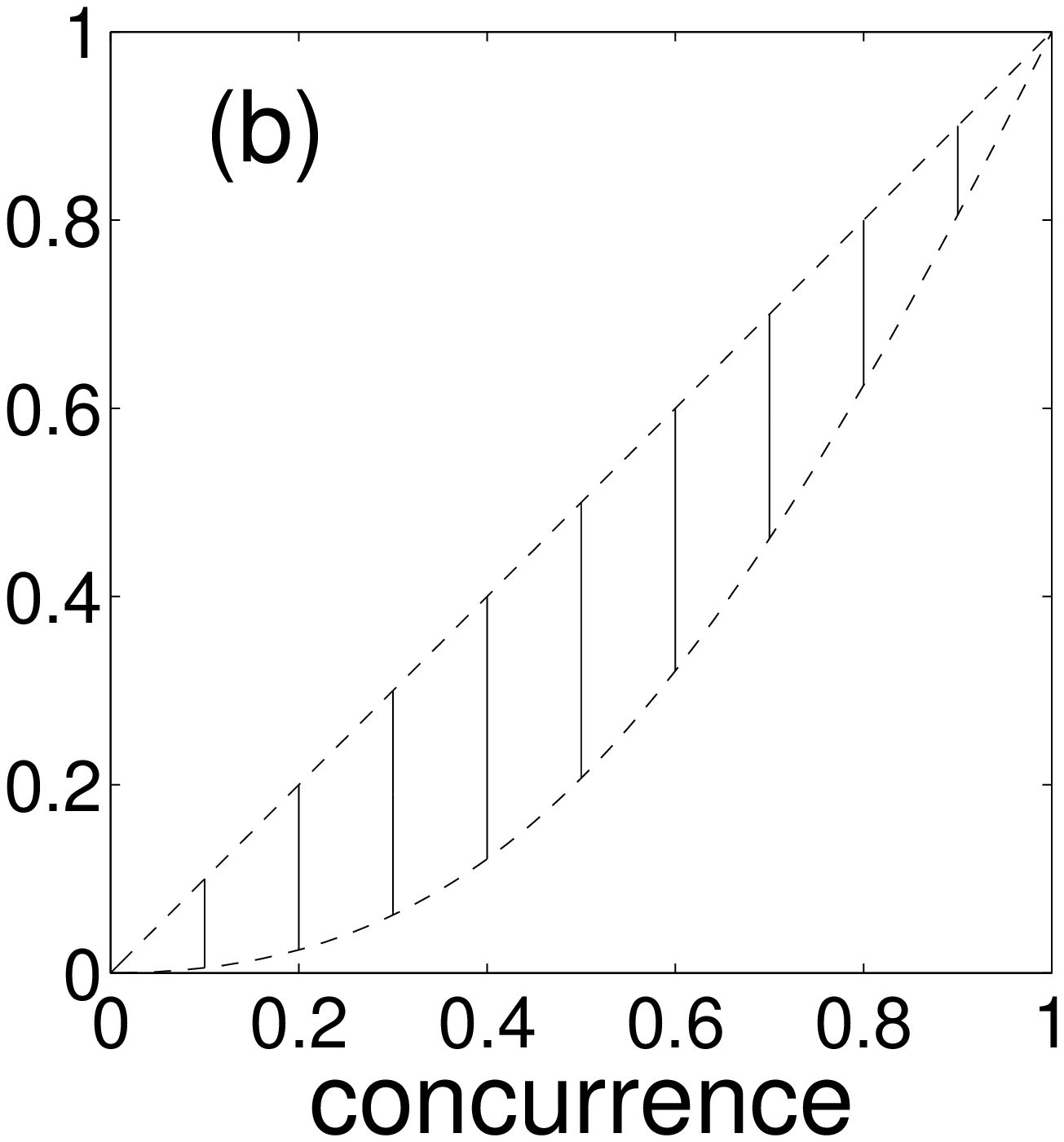}\epsfxsize=2.8cm\epsfbox{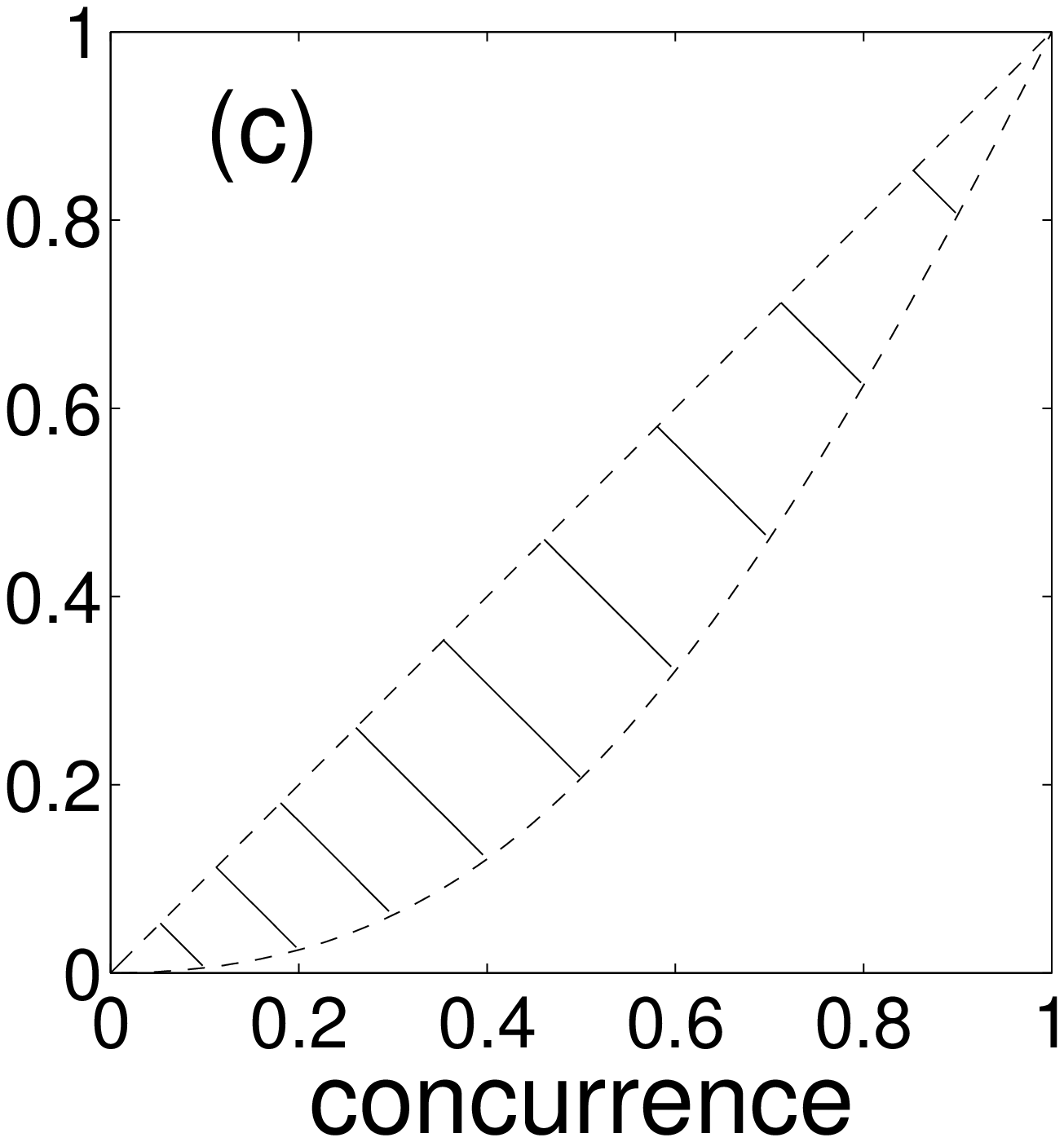}}
\caption{Negativity versus concurrence for the three classes of
states (\ref{N20}) for the parameter $p=0.1,0.2,...,0.9$: (a)
$\rho'$ where $N'=N(\rho_H)$, (b) $\rho''$, and (c) $\rho'''$ where
$\rho=\rho_H$.}
\end{figure}

By analyzing figure 1 for a given state $\rho_1$ corresponding to
some point $(C(\rho_1),N(\rho_1))$, it is easy to identify all
other states $\rho_2$, corresponding to points
$(C(\rho_2),N(\rho_2))$, which lead to violation of (\ref{N01}).
E.g., the state $\rho_1$ described by point O (X) and the other
states $\rho_2$ corresponding to an arbitrary point in regions OPR
and OST (XYZ) violate condition (\ref{N01}). Clearly, maximal
violation of (\ref{N01}) holds if one of the states (say,
$\rho_{1}$) is the MaxNeg state and the other ($\rho_{2}$) is the
MinNeg state. To analyze the degree of violation of (\ref{N01}), we
will calculate $\Delta C\{\rho_1,\rho_2\} \equiv
C(\rho_1)-C(\rho_2)$, $\Delta N\{\rho_1,\rho_2\} \equiv
N(\rho_1)-N(\rho_2)$, and
\begin{equation}
\delta\{\rho_1,\rho_2\} \equiv - \min(0,\Delta C\{\rho_1,\rho_2\}
\Delta N\{\rho_1,\rho_2\}). \label{N15}
\end{equation}
In figure 2, the function $\delta(\rho_1,\rho_2)$ is plotted vs all
possible values of $C(\rho_{1})$ of the MaxNeg states and
$C(\rho_{2})$ of the MinNeg states. A closer look at (\ref{N09}),
leads us to conclusions
\begin{eqnarray}
\max_{{\rho_1,\rho_2} \atop {N(\rho_1)= N(\rho_2)}}\!\!\! |\Delta
C\{\rho_1,\rho_2\}| &=&  \Delta
C\{\rho_X,\rho_Y\}=1-\frac{\sqrt{2}}{2},\quad
\label{N16} \\
\max_{{\rho_1,\rho_2} \atop {C(\rho_1)=C(\rho_2)}} \!\!\! |\Delta
N\{\rho_1,\rho_2\}| &=& \Delta
N\{\rho_Z,\rho_X\}=1-\frac{\sqrt{2}}{2},\quad
\label{N17} \\
\max_{{\rho_1,\rho_2}} \delta\{\rho_1,\rho_2\} &=&
\delta\{\rho_V,\rho_X\}=\frac{\kappa^2}{2}\quad  \label{N18}
\end{eqnarray}
where $\kappa=(\sqrt{2}-1)/2$, $\rho_J$ ($J=V,Y,Z$) are the MaxNeg
states and $\rho_X$ is the MinNeg state having the following
concurrences and negativities: $C(\rho_V)=N(\rho_V)=\sqrt{2}/4$,
$C(\rho_Z)=N(\rho_Z)=C(\rho_X)=1/2$, and
$C(\rho_Y)=N(\rho_Y)=N(\rho_X)=\kappa$, as depicted by the
corresponding points in figure 1. As an explicit example of states
maximally violating (\ref{N01}), one can choose
$\rho_X=\rho_H(1/2)$, given by (\ref{N13}), and the following
Werner states (\ref{N10}): $\rho_Y=\rho_W(\sqrt{2}/3)$,
$\rho_Z=\rho_W(2/3)$, and $\rho_V=\rho_W(1/3+\sqrt{2}/6)$.
Alternatively, instead of the Werner states, one can take the
following pure states (\ref{N07}):
\begin{equation}
|\Psi(p) \rangle =\sqrt{p}\, |01\rangle +\sqrt{1-p}\,|10\rangle
\label{N19}
\end{equation}
which implies that $\rho_Y$ can be given by
$|\Psi(p)\rangle\langle\Psi(p)|$ for $p=1/2 \pm \sqrt{1 +
2\sqrt{2}}/4$; $\rho_Z$ for $p=1/2\pm\sqrt{3}/4$, and $\rho_V$ for
$p=1/2 \pm \sqrt{14}/8$.

Let us also consider the following two-qubit states
\begin{equation}
\bar{\rho}(p,q) = p |\psi_B\rangle \langle \psi_B|+(1-p)
|\psi_q\rangle \langle \psi_q| \label{N20}
\end{equation}
being a mixture of the Bell state, given by (\ref{N11}), and the
separable state
\begin{equation}
|\psi_q\rangle = \sqrt{1-q} |00\rangle + \sqrt{q} |01\rangle
\label{N21}
\end{equation}
where the parameters $p,q\in\langle 0,1\rangle$. The negativity of
$\bar{\rho}(p,q)$ depends on both $p$ and $q$ according to
\begin{equation}
N(\bar{\rho}(p,q)) = \sqrt{1-2p(1-p)(1-q)}-(1-p)\label{N22}
\end{equation}
while the concurrence, given by
\begin{equation}
C(\bar{\rho}(p,q)) = p, \label{N23}
\end{equation}
is clearly independent of $q$. In a special case for $q=0$, Eq.
(\ref{N20}) goes over into Eq. (\ref{N13}) describing the MinNeg
state, while for $q=1$, (\ref{N20}) describes the MaxNeg state as
$N(p,1)=C(p,1)$.

In the following, we will analyze three classes of the states given
by (\ref{N20}). (i) First class is formed by those states with the
same negativity, say $N'$. From Eq. (\ref{N22}), one finds that the
states $\rho'=\bar{\rho}(p,q')$, given by (\ref{N20}) for
\begin{equation}
q'=\frac{N'[N'+2(1-p)]-p^2}{2p(1-p)},
 \label{N24}
\end{equation}
have the $p$-dependent concurrence, $C(\rho')=p$, but constant
negativity, $N(\rho')=N'=$ const, for all $p\in\langle
N',\sqrt{2N'(N'+1)}-N'\rangle$. This result is confirmed
graphically in figure 3(a) for a few choices of $N'$. In
particular, for $N'=\kappa$, one gets the $p$-parametrized
($\kappa\le p\le 1/2$) states
\begin{equation}
\rho_{\rm XY}(p)\equiv \bar{\rho}\left(p,\frac{(1-2p)(2\sqrt{2}
+2p-1)} {8p(1-p)}\right) \label{N25}
\end{equation}
which are visualized by the points on the XY line in figure 1. The
states $\rho_{\rm XY}(p)$ for $p=\kappa$ (corresponding to a MaxNeg
state) and $p=1/2$ (MinNeg state) have maximally different
concurrences and the same negativity,
\begin{eqnarray}
\Delta C\{\rho_{\rm XY}(1/2),\rho_{\rm
XY}(\kappa)\}=1-\frac{\sqrt{2}}{2}. \label{N26}
\end{eqnarray}
(ii) To the second class belong those $\rho''=\bar{\rho}(p'',q)$
having the same concurrence. This condition is easily fulfilled by
fixing $p=p''$ in (\ref{N20}), then the concurrence $C''\equiv
C(p'',q)=p''=$const for all values of $q$, while the negativity
$N(p'',q)$ ranges from $\sqrt{(1-p'')^2+(p'')^2}-(1-p'')$ to $p''$
as shown in figure 3(b) for a few choices of $p''$. In particular,
for $p''=1/2$, we get the $q$-parametrized ($q\in \langle 0,1
\rangle$) states
\begin{equation}
\rho_{\rm XZ}(q)\equiv \bar{\rho}(1/2,q)=\frac12(|\phi\rangle
\langle \phi|+ |\psi_q\rangle \langle \psi_q|) \label{N27}
\end{equation}
which are described by the points on the XZ line in figure 1. The
states $\rho_{\rm XZ}(q)$ for $q=0$ (corresponding to the MinNeg
state) and $q=1$ (MaxNeg state) have maximally different
negativities for the same concurrence,
\begin{eqnarray}
\Delta N\{\rho_{\rm XZ}(1),\rho_{\rm XZ}(0)\}=1-\frac{\sqrt{2}}{2}.
\label{N28}
\end{eqnarray}
(iii) Finally, we analyze such states $\rho'''=\bar{\rho}(p,q''')$
of the form (\ref{N20}) for which predictions concerning negativity
and concurrence are exactly opposite to those for a given state
$\rho$, i.e.,
\begin{equation}
\Delta C\{\rho,\rho'''\} = -\Delta N\{\rho,\rho'''\}. \label{N29}
\end{equation}
This condition is fulfilled if the parameter $q'''$ is given by
\begin{equation}
q'''=1+\frac{[N(\rho)+C(\rho)+1-2p]^2-1}{2p(1-p)} \label{N30}
\end{equation}
for $\frac12 [C(\rho)+N(\rho)]\le p \le C(\rho)<1$ as shown in
figure 3(c). In particular, if $N(\rho)=\kappa$ and $C(\rho)=1/2$,
one arrives at the $p$-parametrized ($\sqrt{2}/4\le p \le 1/2$)
states
\begin{equation}
\rho_{\rm XV}(p)\equiv \bar{\rho}\left(p,\frac{(1-2p)(2\sqrt{2}
+1-2p)} {4p(1-p)}\right) \label{N31}
\end{equation}
exhibiting $N(\rho_{\rm XV}(p))=\sqrt{2}/2-p$ and, as usual,
$C(\rho_{\rm XV}(p))=p$, which are described by the points on the
XV line in figure 1. Predictions of the concurrence and negativity
for the states $\rho_{\rm XV}(p)$ with $p=1/2$ (corresponding to
the MinNeg state) and $p=\sqrt{2}/4$ (MaxNeg state) are maximally
different, as given by
\begin{eqnarray}
\delta\{\rho_{\rm XV}(\frac12),\rho_{\rm XV}(\frac{\sqrt{2}}{4})\}
=\frac{\kappa^2}{2}, \label{N32}
\end{eqnarray}
which is the upper bound determined by (\ref{N18}).


We analyzed ordering of density matrices of two qubits with respect
to the bipartite entanglement quantified by the Wootters
concurrence $C$, a measure of the entanglement of formation, and by
the negativity $N$, a measure of the PPT-entanglement cost. We have
presented simple two-qubit states $\rho_1$ and $\rho_2$ (where one
of them can be pure) having the entanglement measures different in
such a way that (i) $N(\rho_{1})=N(\rho_{2})$ but $C(\rho_{1})\neq
C(\rho_{2})$; (ii) $C(\rho_{1})=C(\rho_{2})$ but $N(\rho_{1})\neq
N(\rho_{2})$, or (iii) the concurrence $C(\rho_{1})$ is smaller
than $C(\rho_{2})$ but the negativity $N(\rho_{1})$ is greater than
$N(\rho_{2})$. Using the bounds of Verstraete et al.
\cite{verstraete}, we have also found analytically to what degree
the concurrence and negativity can give different orderings of
two-qubit states.

{\bf Acknowledgments}. The authors thank Jens Eisert,
Pawe\l{}~Horo\-decki, and Karol \.Zyczkowski for stimulating
discussions. A.G. was supported by the Polish State Committee for
Scientific Research under grant No.~0~T00A 003 23.

\end{document}